\documentclass[twocolumn,preprintnumbers,amsmath,nofootinbib,amssymb]{revtex4}
\usepackage{amssymb}
\usepackage{lmodern}
\usepackage[utf8x]{inputenc}
\usepackage[T1]{fontenc}
\usepackage{latexsym}
\usepackage{color}
\usepackage{enumerate}
\usepackage{graphicx}
\usepackage{hyperref}
\usepackage{color}

\def\nn{\nonumber\\}
\newcommand{\f}[2]{\frac{#1}{#2}}
\def\be{\begin{equation}}
\def\ee{\end{equation}}
\def\bea{\begin{eqnarray}}
\def\eea{\end{eqnarray}}
\begin{document}
\title{Effects of Rastall parameter on perturbation of dark sectors of the Universe}
\author{$^1$A. H. Ziaie\footnote{ah.ziaie@maragheh.ac.ir}, $^2$H. Shabani\footnote{h.shabani@phys.usb.ac.ir} and $^1$S. Ghaffari\footnote{sh.ghaffari@maragheh.ac.ir}}
\address{$^1$Research Institute for Astronomy and Astrophysics of Maragha (RIAAM), University of Maragheh, P.O.Box 55136-553, Maragheh, Iran\\}
\address{$^2${Faculty of Sciences, University of Sistan and Baluchestan, Zahedan, Iran}}
\begin{abstract}
In recent years, Rastall gravity is undergoing a considerable surge in popularity. This theory purports to be a modified gravity theory with a non-conserved energy-momentum tensor ({\rm EMT}) and an unusual non-minimal coupling between matter and geometry. The present work looks for the evolution of homogeneous spherical perturbations within the Universe in the context of Rastall gravity. Using the spherical Top-Hat collapse model we seek for exact solutions in linear regime for density contrast of dark matter (\rm DM) and dark energy ({\rm DE}). We find that the Rastall parameter affects crucially the dynamics of density contrasts for {\rm DM} and {\rm DE} and the fate of spherical collapse is different in comparison to the case of general relativity ({\rm GR}). Numerical solutions for perturbation equations in non-linear regime reveal that {\rm DE} perturbations could amplify the rate of growth of {\rm DM} perturbations depending on the values of Rastall parameter. 
\end{abstract}

\maketitle
\section{Introduction}
The Rastall theory of gravity has been firstly proposed in 1972~\cite{rastall1972}
and then in~\cite{rastall1976} which suggests that 
one may relax conservation of the {\rm EMT}, i.e., it 
assumes $\nabla_\nu T^{\mu\nu}\neq0$ for energy sources~\cite{rastall1972,rastall1976,FESRASPLB,Effemtras,moradpour20171}. The core 
motivation for this choice is that the conservation laws have been only tested on 
the Minkowski spacetime or quasistatic gravitational fields~\cite{moradpour20171}. 
Therefore, one can still modify this presumption accounting for non-zero curvature of spacetime. Another justification for using the theories which violate the EMT conservation is to provide a room to discuss the process of particle production. It is known 
that the conservation of EMT does not lead to the particle creation~\cite{gibbons1977,birrell1982}.
In the Rastall picture, the conservation 
law is proportional to the gradient of the Ricci scalar, i.e., $\nabla_\nu  T^{\mu\nu}= \lambda \nabla^\mu R$. 
Such a relation can be introduced because of 
quantum effects\footnote{Since, in the Rastall gravity the Ricci scalar is related to the trace of energy 
momentum tensor, one may classify this theory as a particular form of $f(R,T)$ 
gravity (see, e.g., Ref.~\cite{shabani2018}).
In $f(R,T)$ theories the implementation of $T$ can be justified by quantum 
effects\cite{shabani2013,shabani2014}.}\cite{fabris2015}. 
It is shown that quantum effects may result in a ``gravitational anomaly" for which the usual conservation
of EMT gets violated~\cite{birrell1982,bertleman1996}. 
Such a phenomenon may have an important impact on black hole Hawking radiation~\cite{fabris2012}.
In the past decades many attempts have been performed to explore the physical contents of the Rastall gravity.
Cosmological consequences of the theory have been discussed in~\cite{batista2010,capone20101,fabris2011,
batista2012,batista2013,silva2013,moradpour20161,moradpour20172}, G\"{o}del-type solutions 
have been investigated in~\cite{santos2015}, the Brans-Dicke scalar field has been considered in 
the Rastall background~\cite{thiago2014,salako2016} and static spherically symmetric solutions have been introduced in~\cite{oliveira2015,oliveira2016,moradpour2016,bronnikov2017,heydarzade2017,lin2019}. The
Rastall proposal has been inspected from Mach's principle landscape however, it is shown that it is compatible with
this principle~\cite{majernik2006}. Moreover, in~\cite{visser2018} it is argued that Rastall gravity is equivalent to {\rm GR} but as discussed in~\cite{darabi2018}, these two theories are not equivalent. Indeed, Rastall theory provides a setting in which geometry and energy-momentum sources can be coupled to each other in a non-minimal way~\cite{moradpour20171,darabi2018,4indexmorad}. Such a mutual interaction can lead to many interesting consequences, see e.g.,~\cite{moradpour20172,darabi2018,Effemtras1} for observational aspects and  ~\cite{yuan2016,moradpour2017,moradpour20177,lobo2018,kumar2018,bamba2018,moradpour2019,halder2019,khyllep2019,mnras2019ras,neutronras,neutronras1,CaiMiao2020,Mota:2019zln,Lobo:2020jfl,Llibre:2020odw,Xi:2020lbt,SinghMishra2020,Mustafa:2020seu,Ziaie:2020ord,Mustafa:2020kng,Prihadi:2019isb,Yu:2019cku,Ziaie:2019klz} for other aspects of the the theory. Specially, in~\cite{ziaie2019} it is shown that the outcome of gravitational collapse in Rastall gravity and {\rm GR} is remarkably different in such a way that, a homogeneous dust collapse in {\rm GR} necessarily leads to black hole formation while either naked singularities and black holes can form as the collapse endstate in Rastall gravity.
Moreover, from cosmological perspective, the differences between Rastall gravity and GR have been reported in~\cite{moradpour20172} and earlier results of~\cite{Smalley_1983} allude to possible non-equivalence of these two theories. Finally, influences of Rastall coupling parameter on modeling of static and spherically symmetric distributions of perfect fluid matter have been surveyed in~\cite{Hansraj_2019} where the authors investigated properties of the well-known stellar model proposed by Tolman~\cite{PhysRev.55.364}. They showed that in most of the case studies, Rastall theory remains well consistent with the basic requirements for physical reliability of the model while the GR theory exhibits defective behavior. The findings of this study and other works~\cite{oliveira2015,Abbas_2018,ABBAS20201} show that Rastall coupling parameter can act as a mathematical tool to compensate the shortcomings of the standard GR. 

%%%%%%%%%%%%%%%%%%%%
\par
The primordial collapsed regions serve as the initial cosmic 
seeds from which the large scale structures like galaxies, clusters, supernovae, quasars etc., are 
developed~\cite{white1978,peebles1993,peacock 1999}. Investigation of the {\rm DE} perturbations in the linear and non-linear level is of great importance.
In these cases, {\rm DE} perturbations may form halo structures influencing the (dark) matter collapsed
region non-linearly~\cite{abramo2007}. Note that, different {\rm DE} scenarios may lead to the same background
expansion rate, nevertheless, they behave differently in the perturbation level. Study of the mutual interactions between
dark sectors in the perturbation level helps us to understand nature of the {\rm DE}. Influence of the {\rm DE} on
structure formation both at the background level (no fluctuations) and the perturbed level has been investigated in 
various scenarios, see e.g.,~\cite{nunes2005,iberato2006,nunes2006,manera2006,dutta2007,abramo2007,adiabaticic,abramo2009}. 
\par
As some new theories are invented to expand the old ones, there may be some motivations to generalize the original Rastall theory of gravity. In this regard, it has been shown that, though Rastall gravity produces acceptable and suitable predictions and explanations for various gravitational and cosmological phenomena~\cite{darabi2018,moradpour2016,yuan2016,mnras2019ras,farookmpla}, a DE-like source is still needed to describe the current accelerated universe in this framework. For example, a thermodynamic description of Friedmann equations in Rastall gravity indicates that DE problem is also valid in this theory~\cite{moradpour20172} and in~\cite{batista2012} it is shown that Rastall theory is consistent with current observations on the Universe expansion whenever a {\rm DE} fluid along with a pressure-less fluid fill the background. This compatibility is seen at background as well as linear perturbation levels, and furthermore, the {\rm DE} candidate may also cluster under the shadow of the Rastall non-minimal coupling between the geometry and cosmic fluids~\cite{batista2012}. Recently, some efforts have been made in order to address the issue of DE through generalizing the original Rastall theory. Such an extension of Rastall gravity has been proposed in~\cite{moradpour20171} where a variable Rastall parameter is utilized instead of the usual constant one ($\lambda$) which is introduced in the original theory. More exactly, the authors of~\cite{moradpour20171} have modified the non-conservation equation as $\nabla_\nu T^{\mu\nu}=\nabla^\mu(\lambda' R)$, where $\lambda'$ is a dynamic variable. This modification reasonably allows a smooth variation of coupling between energy momentum source and geometry that can act as a DE source, responsible for the current accelerating expansion phase. In~\cite{KaiLiang2020}, the authors have investigated a more general case. They assumed a second rank tensor field which is proportional to spacetime metric and a function on Ricci scalar and the trace of EMT. The non-vanishing covariant derivative of this tensor field is considered as the non-conserved sector of the EMT and consequently bears the role of DE during the cosmic evolution. The authors have also found that the amount of violation of EMT is more significant during the DE dominated epoch. More interestingly, the solutions obtained in~\cite{KaiLiang2020} mimic quintessence and k-essence scenarios, which confirm the duality between k-essence theory and Rastall gravity~\cite{Bronnikov2017}.
\par
Hence, in light of the above considerations, investigating the effects of {\rm DE} fluctuations on matter clustering in Rastall gravity can be of interest. A rather simple way to deal with this issue is to utilize the Top-Hat Spherical Collapse ({\rm SC}) model. This approach was initially employed in Einstein-de Sitter background in the standard Cold-{\rm DM} scenario~\cite{tophatsdm}, and later in $\Lambda{\rm CDM}$ model~\cite{tophatLambdacdm}. Work along this line has been also extended to the study of quintessence fields~\cite{tophatqfields}, decaying vacuum models~\cite{tophatdecayvac}, $f(R)$ gravity theories~\cite{tophatfR}, {\rm DE} models with constant equation of state ({\rm EoS})~\cite{abramo2007,tophatdeconseos} and coupled {\rm DE} models~\cite{tophatcoupledde}.
\par Our aim in the present work is to study the evolution of DM and DE perturbations in the framework of the Rastall gravity. 
The paper is then organized as follows. In Sec.~\ref{fesras} we briefly review the field equations of the Rastall 
gravity. In Sec.~\ref{maineqs}, the main evolutionary equations of {\rm DE} and {\rm DM} perturbations 
are obtained. In Sec.~\ref{sollineareqs}, we discuss the linear behavior of fluctuations in matter as well as {\rm DE} dominated eras. Sec.~\ref{nonlinearreg} is devoted to inspecting the non-linear effects and finally 
in Sec.~\ref{conclrems} we summarize our results.
\section{Field equations of Rastall gravity}\label{fesras}
According to the original idea of Rastall, the divergence of {\rm EMT} is proportional to the covariant derivative of Ricci curvature scalar as
\be\label{CoVDiv}
\nabla_{\mu}T^{\mu}_{\,\,\,\nu}=\lambda\nabla_{\nu} R,
\ee
where $\lambda$ is the Rastall parameter. The Rastall field equations are then given by~\cite{FESRASPLB,Effemtras}
\be\label{RastallFES}
G_{\mu\nu} +\gamma g_{\mu\nu}R=\kappa T_{\mu\nu},
\ee
where $\gamma=\kappa\lambda$ is the Rastall dimensionless parameter and $\kappa$ being the Rastall gravitational coupling constant. The above equation can be rewritten in an equivalent form as
\be\label{FESEquiv}
G_{\mu\nu}=\kappa T^{\rm eff}_{\mu\nu},~~~~~T^{\rm eff}_{\mu\nu}=T_{\mu\nu}-\f{\gamma T}{4\gamma-1}g_{\mu\nu},
\ee
where $T^{\rm eff}_{\mu\nu}$ is the effective energy momentum tensor whose components are given by~\cite{rastall1972,moradpour20177}
\bea
T^{0\rm\, eff}_{\,\,\,0}\equiv-\rho^{\rm eff}=-\f{(3\gamma-1)\rho+\gamma(p_r+2p_t)}{4\gamma-1},\label{EFFEMT1}\\
T^{1\rm\, eff}_{\,\,1}\equiv p_r^{\rm eff}=\f{(3\gamma-1)p_r+\gamma(\rho-2p_t)}{4\gamma-1},\label{EFFEMT2}\\
T^{2\rm\, eff}_{\,\,2}=T^{3\rm\, eff}_{\,\,3}\equiv p_t^{\rm eff}=\f{(2\gamma-1)p_t+\gamma(\rho-p_r)}{4\gamma-1}.\label{EFFEMT3}
\eea
We note that in the limit of $\lambda\rightarrow0$ the standard {\rm GR} is recovered. Moreover for an electromagnetic field source we get $T^{\rm eff}_{\mu\nu}=T_{\mu\nu}$ leading to $G_{\mu\nu}=\kappa T_{\mu\nu}$. Therefore, the GR solutions for $T=0$, or equivalently $R=0$, are also valid in the Rastall gravity~\cite{FESRASPLB,bronnikov2017}.
\par
Since the advent of Rastall gravity, there has been a serious curiosity about a more fundamental origin for the Rastall equation (\ref{RastallFES}). As the violation of classical conservation law of the {\rm EMT} is not specific to Rastall
gravitational theory and there are other modified gravity theories that possess such a feature such as $f(R,T)$ gravity~\cite{Harko2011} and $f(R,{\mathcal L}_m)$ gravity~\cite{Harko2010,Harko2014}, one may be motivated to construct a possible Lagrangian formalism to Rastall gravity based on these models. Work along this line has been carried out in~\cite{moraes2019} where the authors have considered different curvature-matter gravity Lagrangians from which the Rastall field equations can be extracted. Two of the present authors have also shown that the Rastall equations can be resulted from $f(R,T)$ gravity Lagrangian, under some conditions~\cite{shabani2020}. It is also noteworthy to consider the geometrical equivalence between Rastall gravity and unimodular (trace-free) theory. The trace of equation (\ref{RastallFES}) is given by	
\bea
(4\gamma-1)R=\kappa T.
\eea
Choosing then $\gamma=1/4$ leads to the result $T=0$ for non-vanishing value of the Ricci scalar (in this case, the Rastall gravity would be applied in the presence of a trace-less energy momentum tensor e.g., a radiation fluid). Substituting $\gamma=1/4$ within equation (\ref{RastallFES}) leaves us with the trace-less form of the Rastall field equations, as follows
\bea
R_{\mu\nu} -\f{1}{4} g_{\mu\nu}R=\kappa T_{\mu\nu}.
\eea
Comparing this result with the trace-less unimodular field equations~\cite{sago,bij}
\bea
R_{\mu\nu} -\f{1}{4} g_{\mu\nu}R= \kappa \left(T_{\mu\nu}-\f{1}{4} g_{\mu\nu}T\right)\nonumber,
\eea
implies that one can find a structural similarity between the Rastall field equations (for particular case $\gamma=1/4$) and those of unimodular theory~\cite{mik,oli,mah,aka}.
\section{Spherical collapse}\label{maineqs}
For a spatially flat, homogeneous and isotropic Universe filled with DM and DE, Eq. (\ref{FESEquiv}) can be put into the form 
\bea
H^2&=&\f{8\pi G}{3(6\gamma-1)}\sum_{k}\left[3\gamma(1+w_k)-1\right]\rho_k\nn&=&\f{8\pi G}{3(6\gamma-1)}\left[(3\gamma-1)(\rho_{\rm m}+\rho_{\rm de})+3\gamma w_{\rm de}\rho_{\rm de}\right],\label{fieldback}\\
\f{\ddot{a}}{a}&=&-\f{4\pi G}{3(6\gamma-1)}\sum_{k}\left[3(2\gamma-1)w_k+(6\gamma-1)\right]\rho_k\nn&=&-\f{4\pi G}{3(6\gamma-1)}\left[3(2\gamma-1)w_{\rm de}\rho_{\rm de}+(6\gamma-1)(\rho_{\rm m}+\rho_{\rm de})\right],\label{fieldback1}\nn
\eea
where, an over-dot denotes derivative with respect to time, $\kappa=2(4\gamma-1)\kappa_G/(6\gamma-1)$, $\kappa_G=4\pi G$, $k=\{{\rm m},{\rm de}\}$ labels {\rm DM} and {\rm DE} components, $H=\dot{a}/a$ is the Hubble parameter, $w_{\rm de}=p_{\rm de}/\rho_{\rm de}$ is the {\rm EoS} parameter of {\rm DE} and $\rho_{\rm m}$, $\rho_{\rm de}$ and $p_{\rm de}$ are the (background) energy densities of {\rm DM} and {\rm DE} and the pressure of {\rm DE}, respectively. The Bianchi identity for Eq. (\ref{FESEquiv}) leaves us with the following continuity equation in Rastall gravity, as
\be\label{conteqH}
\dot{\rho}_j+3H\beta_j\rho_j=0,~~~\beta_j=\beta(\gamma,w_j)=\left[\f{(1+w_j)(4\gamma-1)}{3\gamma(1+w_j)-1}\right].
\ee
This equation describes the density evolution of a single perfect fluid labeled by $j$ with background density $\rho_j$ and pressure $p_j=w_j\rho_j$. Consider now a spherically symmetric region of radius $r$ filled with a homogeneous density $\rho^{{\rm c}}_{j}$ (a top-hat distribution). The {\rm SC} model describes a spherical region with a top-hat profile and uniform density so that at time $t$, $\rho^{{\rm c}}_{j}(t)=\rho_{j}(t)+\delta\rho_{j}$. This region initially undergoes a small perturbation of the background fluid density, i.e., $\delta\rho_{j}$ and is immersed within a homogeneous Universe with energy density $\rho_{j}$. If $\delta\rho_{j}>0$ the spherical region will finally collapse under its own gravitational attraction, otherwise, it will expand faster than the average Hubble flow, generating thus, what is known as a void. Similar to Eq. (\ref{conteqH}), the continuity equation for spherical region can be written as the following form, but now with different {\rm EoS}, i.e., $p^{{\rm c}}_{j}=w^{{\rm c}}_{j}\rho^{{\rm c}}_{j}$  
\be\label{conteqh}
\dot{\rho}^{{\rm c}}_{j}+3h\beta^{{\rm c}}_{j}\rho^{{\rm c}}_{j}=0,~~~\beta^{{\rm c}}_{j}=\beta(\gamma,w^{{\rm c}}_{j})=\left[\f{(1+w^{{\rm c}}_{j})(4\gamma-1)}{3\gamma(1+w^{{\rm c}}_{j})-1}\right],
\ee
where, $h=\dot{r}/r$ denotes the local expansion rate inside the spherical perturbed region and $w^{{\rm c}}_{j}$ denotes the {\rm EoS} in this region. The Friedmann equations for spherical region take the form
\bea
h^2&=&\f{8\pi G}{3(6\gamma-1)}\sum_{k}\left[3\gamma(1+w^{{\rm c}}_k)-1\right]\rho^{{\rm c}}_k\nn&=&\f{8\pi G}{3(6\gamma-1)}\left[(3\gamma-1)(\rho^{{\rm c}}_{\rm m}+\rho^{{\rm c}}_{\rm de})+3\gamma w^{{\rm c}}_{\rm de}\rho^{{\rm c}}_{\rm de}\right],\label{fieldsphreg1}\\
\f{\ddot{r}}{r}&=&-\f{4\pi G}{3(6\gamma-1)}\sum_{k}\left[3(2\gamma-1)w^{{\rm c}}_{k}+(6\gamma-1)\right]\rho^{{\rm c}}_{k}\nn&=&-\f{4\pi G}{3(6\gamma-1)}\left[3(2\gamma-1)w^{\rm c}_{\rm de}\rho^{\rm c}_{\rm de}+(6\gamma-1)(\rho^{\rm c}_{\rm m}+\rho^{\rm c}_{\rm de})\right],\label{fieldsphreg2}\nn
\eea
where the second equation governs the dynamics of radius $r$ of the collapsing region. We note, in general, that the densities and pressures obey different {\rm EoS}s inside and outside the spherical region, i.e., $w^{{\rm c}}_{j}\neq w_{j}$. Indeed, the difference between the local and background {\rm EoS}s, $\delta w_j\equiv w^{{\rm c}}_{j}-w_{j}$ can be related to the effective sound speed of the fluid, ${\sf C}^2_{{\rm eff}j}=\delta p_j/\delta\rho_j$. This relation can be re-expressed through introducing the density contrast of a single fluid species labeled by $j$ 
\be\label{dencont}
\delta_j=\f{\rho^{\rm c}_j}{\rho_j}-1=\f{\delta\rho_j}{\rho_j}.
\ee
We therefore have
\be\label{wjwcj}
w^{{\rm c}}_{j}=\f{p^{{\rm c}}_{j}}{\rho^{{\rm c}}_{j}}=\f{p_j+\delta p_j}{\rho_j+\delta\rho_j}=w_j+\left({\sf C}^2_{{\rm eff}j}-w_j\right)\f{\delta_j}{1+\delta_j}.
\ee
The above equation provides a relation between {\rm EoS} within the perturbed region and that of the background, the effective sound speed and the size of perturbations. In the present model, we consider the case in which the {\rm EoS}s inside the collapsing region and the background are identical. We therefore take $\delta w_j=0$ leading to ${\sf C}^2_{{\rm eff}j}=w_j$ and $\beta^{{\rm c}}_{j}=\beta_{j}$. Differentiating Eq. (\ref{dencont}) with respect to time gives
\be\label{dencontdt}
\dot{\delta}_j=3(1+\delta_j)(H-h)\beta_j,
\ee
where we have used Eqs. (\ref{conteqH}) and (\ref{conteqh}). Differentiating again with respect to time leaves us with the following equation for amplitude of the perturbations
\bea\label{dencontdtdt}
\ddot{\delta}_j&=&\left[\dot{w}_j\f{d\ln\beta_j}{dw_j}-2H\right]\dot{\delta}_j\nn&+&\f{4\pi G\beta_j}{6\gamma-1}(1+\delta_j)\sum_{k}\left[3(2\gamma-1)w_k+6\gamma-1\right]\rho_k\delta_k\nn&+&\f{3\beta_j+1}{3\beta_j}\f{\dot{\delta}^2_j}{1+\delta_j},
\eea
where use has been made of Eqs. (\ref{fieldback}), (\ref{fieldback1}), (\ref{fieldsphreg1}), (\ref{fieldsphreg2}) and (\ref{dencontdt}). We note that for $\gamma=0$, we have $\beta_j=1+w_j$ and Eq. (\ref{dencontdtdt}) reduces to its counterpart given in~\cite{abramo2007}. For a mixture of {\rm DM} (here we do not distinguish between {\rm DM} and baryons) and {\rm DE} gravitationally interacting with each other, the top-hat spherical regions evolve according to the following system of differential equations
\bea\label{denconmatter}
\ddot{\delta}_{\rm m}&+&2H\dot{\delta}_{\rm m}-\f{(15\gamma-4)\dot{\delta}^2_{\rm m}}{3(4\gamma-1)(1+\delta_{\rm m})}\nn&=&\f{3H^2(4\gamma-1)}{2(3\gamma-1)(6\gamma-1)}(1+\delta_{\rm m})\Big[(6\gamma-1)\Omega_{\rm m}\delta_{\rm m}\nn
&+&\Big(3(2\gamma-1)w_{{\rm de}}+6\gamma-1\Big)\Omega_{\rm de}\delta_{\rm de}\Big],
\eea
for the density contrast in {\rm DM} component i.e., $\delta_{\rm m}$, and
\bea\label{dencondark}
\ddot{\delta}_{\rm de}&+&\left[2H-\dot{w}_{{\rm de}}\f{d\ln\beta_{{\rm de}}}{dw_{{\rm de}}}\right]\dot{\delta}_{\rm de}\nn&-&\left[\f{3(1+w_{\rm de})(5\gamma-1)-1}{3(1+w_{\rm de})(4\gamma-1)}\right]\f{\dot{\delta}^2_{\rm de}}{1+\delta_{\rm de}}\nn&=&\f{3H^2(1+w_{\rm de})(4\gamma-1)(1+\delta_{\rm de})}{2(6\gamma-1)\Big(3\gamma(1+w_{\rm de})-1\Big)}\Big[(6\gamma-1)\Omega_{\rm m}\delta_{\rm m}\nn&+&\Big(3(2\gamma-1)w_{{\rm de}}+6\gamma-1\Big)\Omega_{\rm de}\delta_{\rm de}\Big],\nn
\eea
for density contrast in {\rm DE} component i.e., $\delta_{\rm de}$. In these equations we have set $w_{\rm m}=0$ and {\rm DM} and {\rm DE} density parameters are defined respectively as
\be\label{denspardmde}
\Omega_{\rm m}=\f{8\pi G \rho_{\rm m}}{3H^2},~~~~~\Omega_{\rm de}=\f{8\pi G \rho_{\rm de}}{3H^2}.
\ee
\section{Solutions in linear regime}\label{sollineareqs}
In order to extract some physical results from Eqs. (\ref{denconmatter}) and (\ref{dencondark}) we rewrite them for constant value of $w_{{\rm de}}$ along with neglecting the terms containing ${\mathcal O}(\delta^2)$. We then have  
\bea
\ddot{\delta}_{\rm m}&+&2H\dot{\delta}_{\rm m}\nn&=&\f{3(4\gamma-1)H^2}{2(3\gamma-1)(6\gamma-1)}\Big[(6\gamma-1)\Omega_{\rm m}\delta_{\rm m}\nn
&+&\Big(3(2\gamma-1)w_{{\rm de}}+6\gamma-1\Big)\Omega_{\rm de}\delta_{\rm de}\Big],\label{denconmatterlin}\\
\ddot{\delta}_{\rm de}&+&2H\dot{\delta}_{\rm de}\nn&=&\f{3(1+w_{\rm de})(4\gamma-1)H^2}{2(6\gamma-1)\Big(3\gamma(1+w_{\rm de})-1\Big)}\Big[(6\gamma-1)\Omega_{\rm m}\delta_{\rm m}\nn&+&\Big(3(2\gamma-1)w_{{\rm de}}+6\gamma-1\Big)\Omega_{\rm de}\delta_{\rm de}\Big].\label{dencondarklin}
\eea
\subsection{Matter dominated era}
\begin{figure}
\includegraphics[scale=0.23]{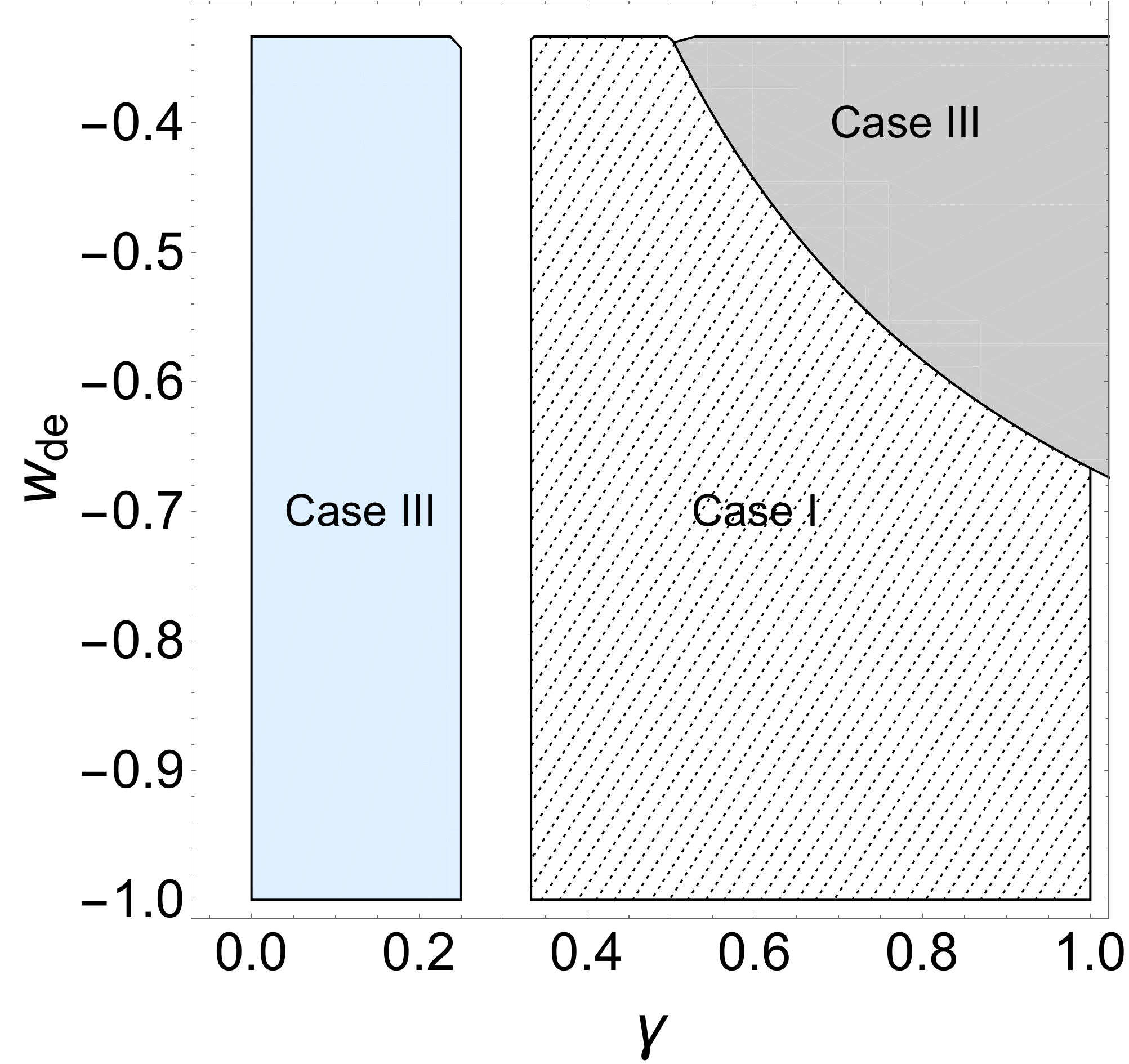}\vspace{0.4cm}
\includegraphics[scale=0.23]{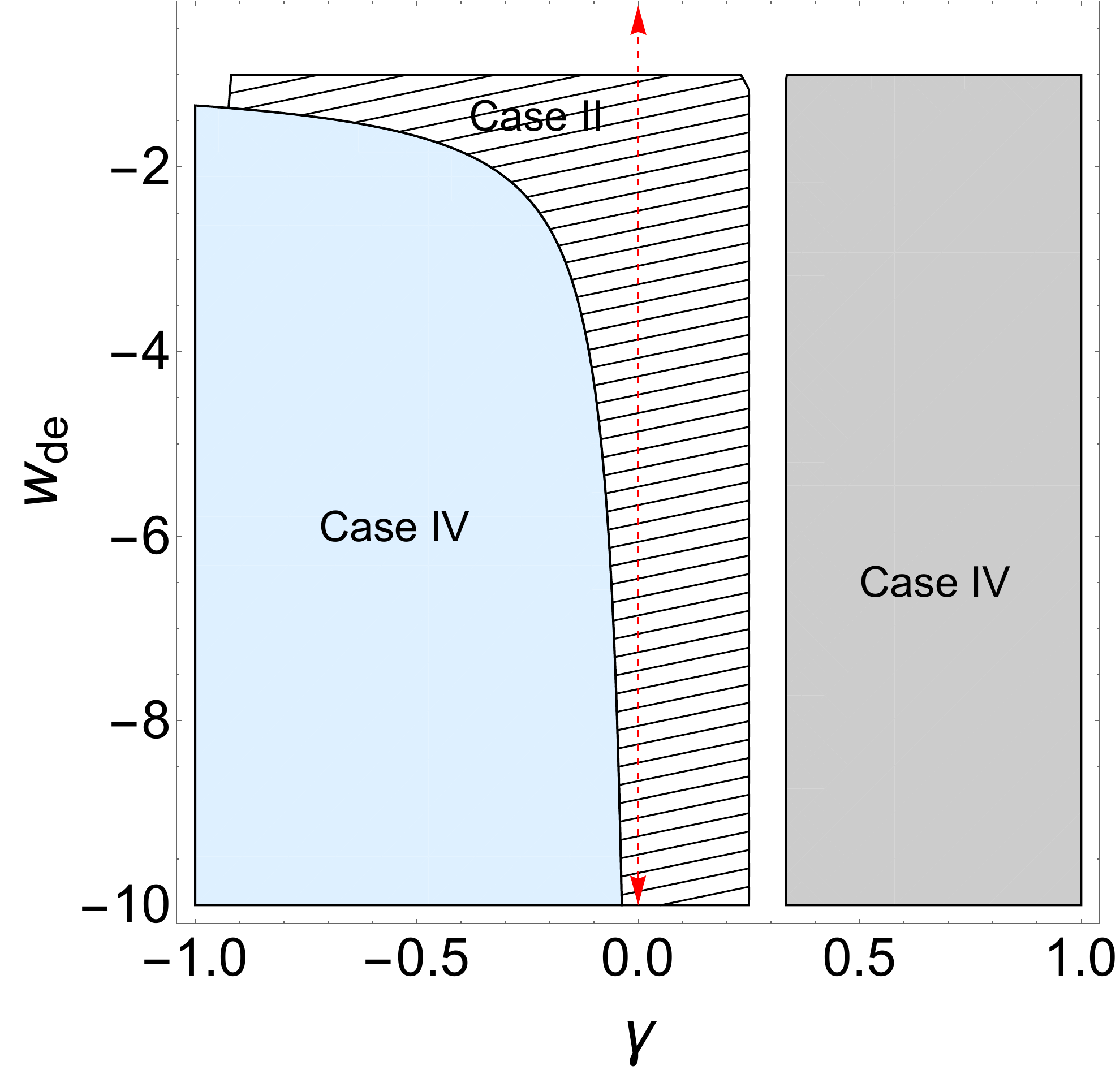}
\caption{The allowed values for {\rm DE} equation of state and Rastall parameters, subject to the conditions given in cases {\rm I-IV}. The red dashed arrow corresponds to $\gamma=0$. The white region is not allowed for the model parameters.}\label{FIGWG1}
\end{figure}

In principle, one can utilize any suitable parameterization for {\rm DE} as a function of time or redshift. However, to obtain analytical solutions, we consider Eqs. (\ref{denconmatterlin}) and (\ref{dencondarklin}) within the matter dominated epoch ($z=10^3$), when the density parameters for {\rm DM} and {\rm DE} can be approximated as $\Omega_{\rm m}\approx1$ and $\Omega_{\rm de}\approx0$, respectively. We then have
\bea
\delta^{\prime\prime}_{\rm m}+\f{3\delta^\prime_{{\rm m}}}{2a}-\f{3(4\gamma-1)}{2(3\gamma-1)a^2}\delta_{{\rm m}}=0\label{deltamprime}\\
\delta^{\prime\prime}_{\rm de}+\f{3\delta^\prime_{{\rm de}}}{2a}-\f{3(1+w_{{\rm de}})(4\gamma-1)}{2a^2\Big(3\gamma(1+w_{\rm de})-1\Big)}\delta_{{\rm m}}=0,\label{deltamprime1}
\eea
where a prime denotes a derivative with respect to $a$ and use has been made of Eqs. (\ref{fieldback1}). It is straightforward to find the analytic solutions of the above equations. We can firstly solve Eq. (\ref{deltamprime}) for matter density contrast with the solution given as
\be\label{deltamprimesol}
\delta_{{\rm m}}(a)={\sf C}_1a^{\alpha_1}+{\sf C}_2a^{\alpha_2},
\ee
where %\alpha_2={\f{1}{4}\left[-1+\sqrt{\f{99\gamma-25}{3\gamma-1}}\right]}
\be\label{alpha12}
\alpha_{1,2}={-\f{1}{4}\left[1\pm\sqrt{\f{99\gamma-25}{3\gamma-1}}\right]},
\ee
and ${\sf C}_1$ and ${\sf C}_2$ are integration constants. We then realize that for those values of Rastall parameter which belong to the set ${\sf S}=\left\{\gamma\mid\gamma\in \mathbb{R}:\gamma<1/4\lor\gamma>1/3\right\}$, we have $\alpha_1<0$ and $\alpha_2>0$, always. Therefore, as the Universe expands, the first term in Eq. (\ref{deltamprimesol}) decays but the second one increases leading to a growing matter density contrast. The matter density contrast decreases for $1/4<\gamma\leq25/99$ as for this case both the exponents of scale factor assume negative values. However, this case cannot be of interest as we are dealing with matter dominated era\footnote{We also note that for $\frac{25}{99}<\gamma <\frac{1}{3}$, the exponents $\alpha_{1,2}$ assume complex values so that solutions including hyperbolic functions, namely $\cosh$ and $\sinh$ will appear. However, we will not deal with these type of solutions in the present study.}. We note that for $\gamma=0$, the solution obtained in~\cite{abramo2007} is recovered. Now, if we neglect the decaying mode in Eq. (\ref{deltamprimesol}) and substitute the result into Eq. (\ref{deltamprime1}), we obtain the following solution for the amplitude of {\rm DE} perturbations as
\be\label{deltamprimesolde}
\delta_{{\rm de}}(a)={\sf C}_3-\f{2{\sf C}_4}{\sqrt{a}}+\xi(\gamma,w_{{\rm de}})\delta_{{\rm m}}(a),
\ee
where 
\be\label{Xgammaw}
\xi(\gamma,w_{{\rm de}})=\f{(3\gamma-1)(1+w_{{\rm de}})}{3\gamma(1+w_{\rm de})-1}.
\ee
At first glance we observe that the evolution of density contrast of {\rm DE} depends on the Rastall parameter as well as the {\rm EoS} of {\rm DE}. It is natural to choose the adiabatic initial condition for {\rm DE} density contrast~\cite{adiabaticic}, i.e., setting ${\sf C}_3=0$. We note that the adiabatic condition is different in the usual {\rm DE} models for which $w_{{\rm de}}>-1$ as compared to phantom models for which $w_{{\rm de}}<-1$. Let us first consider the case $\gamma=0$ for which $\xi(0,w_{{\rm de}})=1+w_{{\rm de}}$. This case has been studied before by the authors of~\cite{abramo2007} and we here give a brief review on it. Neglecting then the decaying term, we see that for phantom models, adiabatic initial conditions mean that, any initial over-density in {\rm DM} ($\delta_{\rm m}>0$) is accompanied by an under-density in {\rm DE} ($\delta_{\rm de}<0$) and vice versa. The case ${\sf C}_3\neq0$ implies a non-adiabatic initial condition, i.e., the perturbations bear an isocurvature component. In this case, if we assume initially positive densities for {\rm DM} and {\rm DE} perturbations, namely $\delta_{{\rm m}}^{\rm i}>0$ and $\delta_{{\rm de}}^{\rm i}>0$, we have 
\be\label{deltamnonad}
\delta_{{\rm de}}(a)=\delta_{{\rm de}}^{\rm i}+(1+w_{{\rm de}})\left(\delta_{{\rm m}}(a)-\delta_{{\rm m}}^{\rm i}\right).
\ee
For $\gamma\in{\sf S}$, we always have a growing density contrast for {\rm DM}, hence $\delta_{{\rm m}}(a)\geq\delta_{{\rm m}}^{\rm i}$. Therefore, if initially the {\rm DE} perturbations are positive, then the pressure gradients, due to non-adiabatic perturbations, will make the {\rm DE} halo to decay, till a critical value for the scale factor, for which $\delta_{{\rm de}}(a_{{\rm cr}})=0$, is reached. This critical value is given by
\be\label{acr}
a_{{\rm cr}}=\f{\delta_{{\rm de}}^{\rm i}}{\mid1+w_{{\rm de}}\mid}+\delta_{{\rm m}}^{\rm i},~~~~~~w_{{\rm de}}<-1.
\ee
We note that for a vanishing Rastall parameter, $\alpha_1=-3/2$ and $\alpha_2=1$, where the former corresponds to a decaying mode and the latter corresponds to a growing mode for {\rm DM} perturbations. For $a>a_{{\rm cr}}$, the {\rm DE} density contrast turns to negative values leaving thus the scenario with a {\rm DE} void~\cite{abramo2007}. Next, we proceed to study Eq. (\ref{deltamprimesolde}) for non-vanishing Rastall parameter. Assuming adiabatic initial conditions (${\sf C}_3=0$) together with neglecting the decaying term, we can deduce the following propositions:

\begin{enumerate}[I\,)]
\item \hspace{0.8cm}$\gamma>\f{1}{3}$~~$\land$~~$\xi(\gamma,w_{{\rm de}})<0$~~$\land$~~$ -1<w_{\rm de}<-\f{1}{3}$,\vspace{0.5cm}\\\vspace{0.5cm}or equivalently \\ $\Big\{\f{1}{3}<\gamma\leq\f{1}{2}\land-1<w_{\rm de}<-\f{1}{3}\Big\}$$\lor$\\$\left\{\gamma>\f{1}{2}\land-1<w_{\rm de}<\f{1-3\gamma}{3\gamma}\right\}$,\vspace{0.5cm}
\item \hspace{0.5cm}$\gamma<\f{1}{4}$~~$\land$~~$\xi(\gamma,w_{{\rm de}})<0$~~$\land$~~$ w_{{\rm de}}<-1$,\vspace{0.5cm}\\\vspace{0.5cm}or equivalently \\$\Big\{\gamma<0\land\f{1-3\gamma}{3\gamma}<w_{\rm de}<-1\Big\}$$\lor$\\$\Big\{0\leq\gamma<\f{1}{4}\land w_{\rm de}<-1\Big\}$,\vspace{0.5cm}
\item \hspace{0.5cm}$\gamma\in{\sf S}$~~$\land$~~$\xi(\gamma,w_{{\rm de}})>0$~~$\land$~~$-1< w_{{\rm de}}<-\f{1}{3}$,\vspace{0.5cm}\\\vspace{0.5cm}or equivalently \\
$\Big\{\gamma>\f{1}{2}\land\f{1-3\gamma}{3\gamma}<w_{\rm de}<-\f{1}{3}\Big\}$~$\lor$\\$\Big\{\gamma<\f{1}{4}\land-1<w_{\rm de}<-\f{1}{3}\Big\}$,
\item \hspace{0.5cm}$\gamma\in{\sf S}$~~$\land$~~$\xi(\gamma,w_{{\rm de}})>0$~~$\land$~~$ w_{{\rm de}}<-1$,\vspace{0.5cm}\\\vspace{0.5cm}or equivalently \\
$\Big\{\gamma>\f{1}{3}\land w_{\rm de}<-1\Big\}$$\lor$$\Big\{\gamma<0\land w_{\rm de}<\f{1-3\gamma}{3\gamma}\Big\}$.
\end{enumerate}
The items {\rm I} and {\rm III} deal with usual {\rm DE} models and those of {\rm II} and {\rm IV} deal with phantom models. The first item corresponds to the case in which an initial over-density in {\rm DM} component is matched by an under-density in {\rm DE} and as the collapse proceeds, voids of {\rm DE} can be formed. However, it is still possible to have an initial over-density for {\rm DE} component with $w_{\rm de}>-1$. As indicated in case {\rm III}, if initially there is an over-density in matter, i.e., $\delta_{{\rm m}}^{\rm i}>0$, we then have $\delta_{{\rm de}}^{\rm i}>0$. This case is the counterpart of case {\rm I} but with different evolution for {\rm DE} perturbations. More interestingly, as shown in case {\rm IV}, an initial overdensity for {\rm DE} can also occur for phantom {\rm DE} models. For this case, any initial overdensity in {\rm DM} leads to an initial overdensity in {\rm DE} component, hence, overdense regions of {\rm DE} would be more and more overdense as time elapses. The case II provides similar situation as reported in~\cite{abramo2007}. In Fig. (\ref{FIGWG1}) we have encapsulated the conditions {\rm I-IV} where the allowed regions for the pair $(w_{\rm de},\gamma)$ are presented.
\par 
For non-adiabatic initial conditions Eq. (\ref{deltamprimesolde}) can be re-expressed as 
\be\label{deltamprimesolderex}
\delta_{{\rm de}}(a)=\delta_{{\rm de}}^{\rm i}+\xi(\gamma,w_{{\rm de}})\left(\delta_{{\rm m}}(a)-\delta_{{\rm m}}^{\rm i}\right).
\ee
Now consider again the cases {\rm I} and {\rm II}. If we take $\delta_{{\rm de}}^{\rm i}>0$ and $\delta_{{\rm m}}^{\rm i}>0$, then the pressure gradients give rise to {\rm DE} decay, turning it into {\rm DE} void through switching the sign of {\rm DE} perturbation at a critical value of the scale factor for which $\delta_{{\rm de}}(a_{\rm cr}^\star)=0$
\be\label{acrstar}
a_{\rm cr}^\star=\left[\delta_{{\rm m}}^{\rm i}+\f{\delta_{{\rm de}}^{\rm i}}{\mid\xi\mid}\right]^{\f{1}{\alpha_2}}.
\ee
This situation can occur for both phantom and usual {\rm DE} models, in comparison to the situation presented in~\cite{abramo2007} which can occur only for phantom models. From Eqs. (\ref{deltamnonad}) and (\ref{acr}) we see that it is only the EoS of DE that decides the evolution of DE perturbations from an initial seed of fluctuations in DM component. However, in Rastall gravity, the mutual matter-geometry interaction (encoded within the $\gamma$ coupling parameter) can provide a different scenario for {\rm DE} perturbations that arise from non-adiabatic initial conditions. We therefore conclude that the presence of Rastall parameter could crucially alter the evolution of {\rm DM} and {\rm DE} perturbations in matter dominated era.
\subsection{Dark Energy Dominated era}
\begin{figure}
	\includegraphics[scale=0.22]{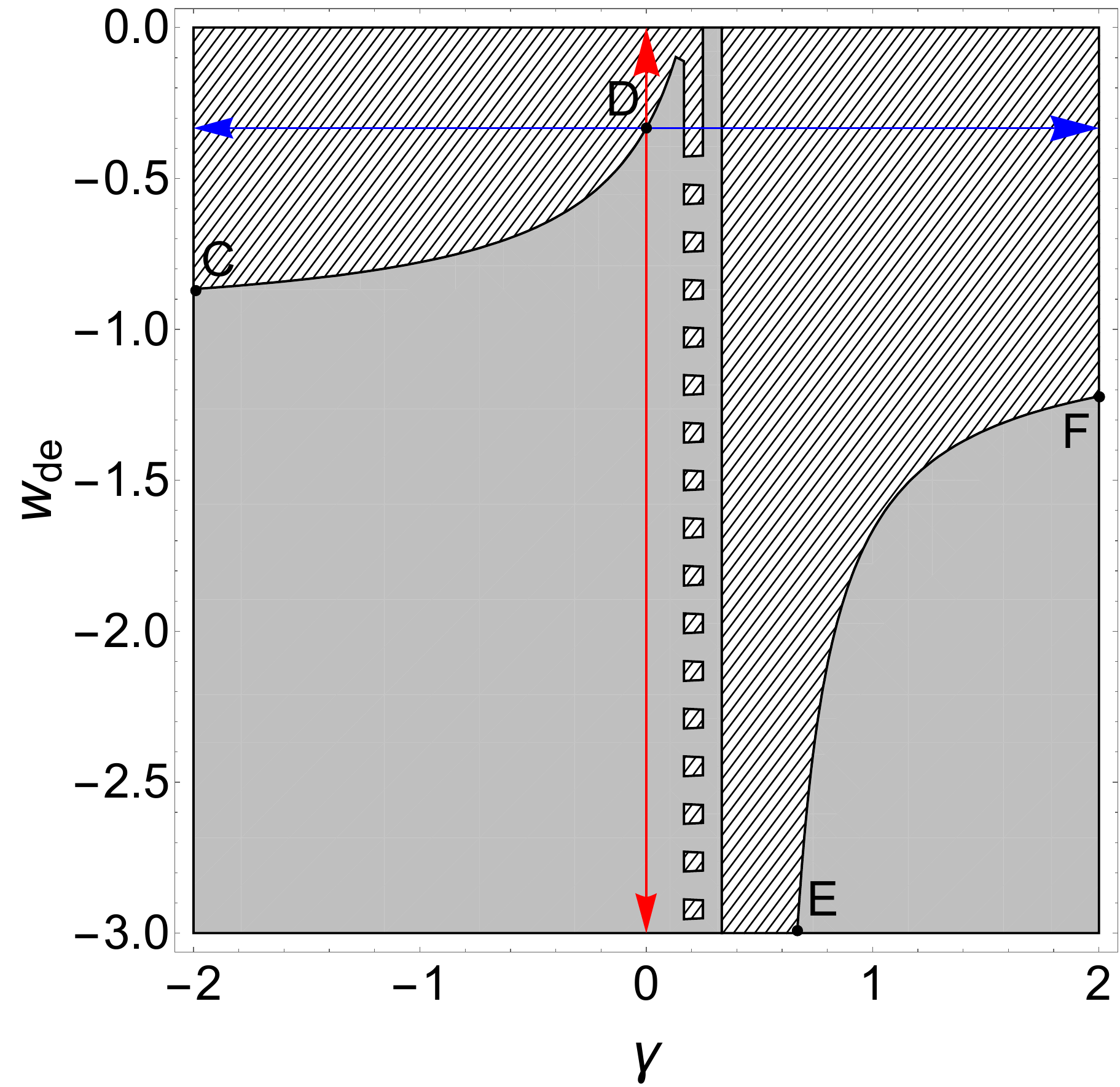}
	\caption{The allowed values for {\rm DE} equation of state and Rastall parameters for which $b_2>0$ (shaded region) and $b_2<0$ (gray region). The red and blue arrows represent $\gamma=0$ and $w_{\rm de}=-1/3$ limits, respectively. The white region is not allowed for the model parameters.}\label{fig2}
\end{figure}

\begin{figure}
	\includegraphics[scale=0.5]{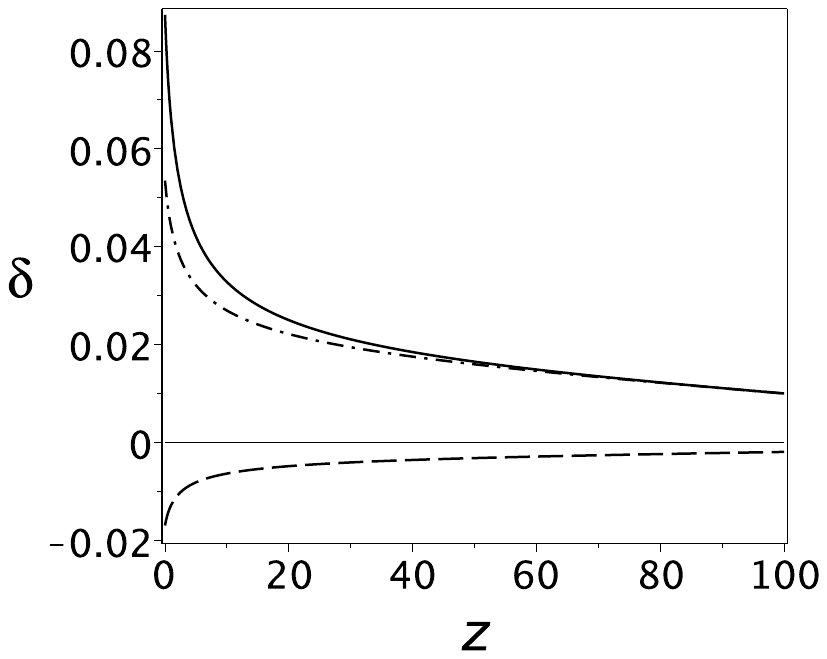}\vspace{-8.5cm}
	\includegraphics[scale=0.5]{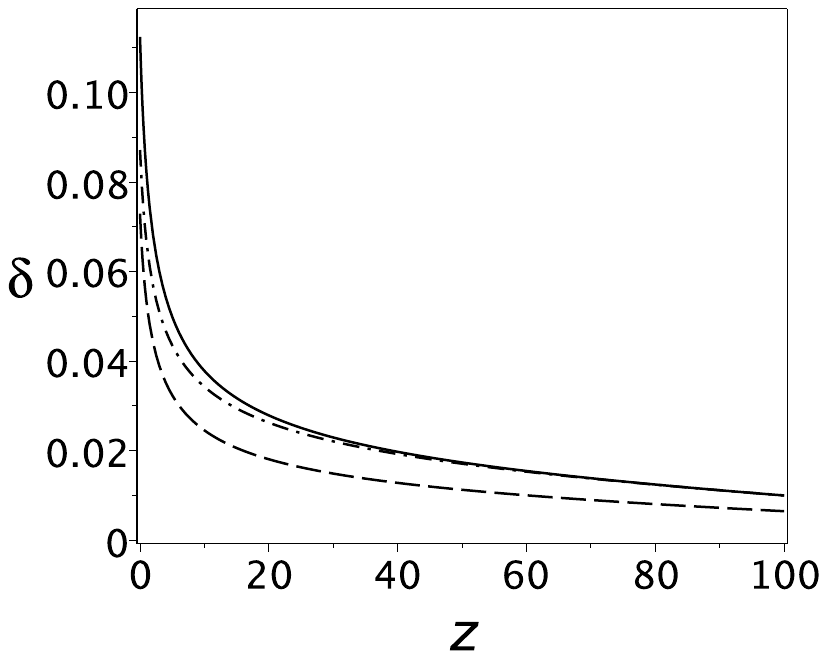}\vspace{-7.5cm}
	\caption{Evolution of density contrasts $\delta_{{\rm m}}$ (solid curve), $\delta_{{\rm de}}$ (dashed curve) and $\tilde{\delta}_{{\rm m}}$ (dot-dashed curve) for $\gamma=-0.263$ and $w_{\rm de}=-1.11$ (upper panel) and $\gamma=-0.975$ and $w_{\rm de}=-0.68$ (lower panel). For upper and lower panels, we have chosen the pair $(\gamma,w_{\rm de})$ from gray and shaded regions of Fig. (\ref{fig2}), respectively. For density parameters we have set $\Omega_{\rm de}=1-\Omega_{\rm m}=3/4.$}\label{fig3}
\end{figure}

In order to study the effects of {\rm DE} perturbations on the evolution of {\rm DM} perturbations, we consider Eqs. (\ref{denconmatterlin}) and (\ref{dencondarklin}) in {\rm DE} dominated period. Let us begin with Eq. (\ref{fieldback1}) which can be put into the following form
\be\label{fieldback1rex}
\ddot{a}=-\f{aH^2}{2(6\gamma-1)}\Big[(6\gamma-1)(\Omega_{\rm m}+\Omega_{\rm de})+3(2\gamma-1)w_{\rm de}\Omega_{\rm de}\Big].
\ee
Now, considering the following transformations for derivatives
\bea\label{transderivatives}
\dot{\delta}=aH\delta^{\prime},~~~~~~~~~~\ddot{\delta}=\ddot{a}\delta^{\prime}+a^2H^2\delta^{\prime\prime},
%\!\!&&\!\!\!\!\!\!\!\!\!\!\!\!\!\!\!\delta^{\prime}=-(1+z)^2\f{d\delta}{dz},~\delta^{\prime\prime}=(1+z)^4\f{d^2\delta}{dz^2}+2(1+z)^3\f{d\delta}{dz},
\eea
together with using Eq. (\ref{fieldback1rex}) we can re-express Eqs. (\ref{denconmatterlin}) and (\ref{dencondarklin}) in terms of scale factor derivatives as
\bea
&&a^2\delta_{\rm m}^{\prime\prime}+ab_0\delta_{\rm m}^{\prime}-b_1\delta_{\rm m}=b_2\delta_{\rm de},\label{deltamdea}\\
&&a^2\delta_{\rm de}^{\prime\prime}+ab_0\delta_{\rm de}^{\prime}-b_4\delta_{\rm de}=b_3\delta_{\rm m},\label{deltamdea1}
\eea
where
\bea\label{ab01234}
b_0&=&2-\f{1}{2}\left[\Omega_{\rm m}+\left(1+\f{3(2\gamma-1)}{6\gamma-1}w_{\rm de}\right)\Omega_{\rm de}\right],\nn
b_1&=&\f{3(4\gamma-1)}{2(3\gamma-1)}\Omega_{\rm m},\nn
b_2&=&\f{3(4\gamma-1)\Big(3(2\gamma-1)w_{\rm de}+6\gamma-1\Big)}{2(3\gamma-1)(6\gamma-1)}\Omega_{\rm de},\nn
b_3&=&\f{3(1+w_{\rm de})(4\gamma-1)}{2\left(3\gamma(1+w_{\rm de})-1\right)}\Omega_{\rm m},\nn
b_4&=&\f{(1+w_{\rm de})(3\gamma-1)b_2}{3\gamma(1+w_{\rm de})-1}.\nn
\eea
The system of differential equations (\ref{deltamdea}) and (\ref{deltamdea1}) admits the following solutions for {\rm DM} and {\rm DE} perturbations as
\bea
\delta_{{\rm m}}(a)&=&{\sf C}_5a^{-\f{1}{2}(b_0+B-1)}+{\sf C}_6a^{\f{1}{2}(B-b_0+1)}\nn&+&{\sf C}_7a^{-\f{1}{2}(b_0+B_1-1)}+{\sf C}_8a^{\f{1}{2}(B_1-b_0+1)},\label{deltamsolution}\\
\delta_{{\rm de}}(a)&=&\f{1}{2b_2}\Big[{\sf C}_5Aa^{-\f{1}{2}(b_0+B-1)}+{\sf C}_6Aa^{\f{1}{2}(B-b_0+1)}\nn&+&{\sf C}_7A_1a^{-\f{1}{2}(b_0+B_1-1)}+{\sf C}_8A_1a^{\f{1}{2}(B_1-b_0+1)}\Big],\nn\label{deltadesolution}
\eea
where
\bea\label{Bsbs}
B\!\!\!&=&\!\!\!\Big[b_0^2-2\sqrt{(b_1-b_4)^2+4b_2b_3}\nn&-&2b_0+2b_1+2b_4+1\Big]^{\f{1}{2}},\nn
B_1\!\!\!&=&\!\!\!\Big[b_0^2+2\sqrt{(b_1-b_4)^2+4b_2b_3}\nn&-&2b_0+2b_1+2b_4+1\Big]^{\f{1}{2}},\nn
A\!\!\!&=&\!\!\!b_4-\sqrt{(b_1-b_4)^2+4b_2b_3}-b_1,\nn
A_1\!\!\!&=&\!\!\!b_4+\sqrt{(b_1-b_4)^2+4b_2b_3}-b_1.
\eea
Note that in the limit of $\gamma\rightarrow0$ we have
\bea\label{gamma-0bs}
&&b_0\rightarrow2-\f{1}{2}\Omega_{\rm m}-\f{1}{2}(1+3w_{\rm de})\Omega_{\rm de},\nn
&&b_1\rightarrow \f{3}{2}\Omega_{\rm m},~~~b_2\rightarrow\f{3}{2}(1+3w_{\rm de})\Omega_{\rm de},\nn
&&b_3\rightarrow\f{3}{2}(1+w_{\rm de})\Omega_{\rm m},~~b_4\rightarrow\f{3}{2}(1+w_{\rm de})(1+3w_{\rm de})\Omega_{\rm de}.\nn
\eea
The unknown constants ${\sf C}_5-{\sf C}_8$ can be determined using the adiabatic initial conditions~\cite{abramo2007,abramo2009}
\bea\label{adiainitialras}
\f{d\delta_{{\rm m}}}{dz}\Big|_{z=z_i}\!\!\!&=&-\f{\alpha_2\delta_{\rm m}(z_i)}{1+z_i},~~\f{d\delta_{{\rm de}}}{dz}\Big|_{z=z_i}\!\!\!=-\f{\alpha_2\xi\delta_{\rm m}(z_i)}{1+z_i},\nn~~~\delta_{{\rm de}}(z_i)&=&\xi\delta_{{\rm m}}(z_i),
\eea
whence we finally obtain solutions (\ref{deltamsolution}) and (\ref{deltadesolution}) in terms of redshift $1+z=a^{-1}$, as
\bea
\delta_{{\rm m}}(z)&=&\f{(1+z_i)^{2(1-b_0)}\delta_{{\rm m}}(z_i)}{BB_1(A_1-A)}\Bigg[B_1(2\xi b_2-A_1)\times\nn&&\!\!\!\!\!\!\!\!\!\!\!\!\Bigg(q_1\f{(1+z)^{\f{1}{2}(b_0+B-1)}}{(1+z_i)^{\f{1}{2}(B-3b_0+3)}}-q_0\f{(1+z)^{\f{1}{2}(b_0-B-1)}}{(1+z_i)^{-\f{1}{2}(B+3b_0-3)}}\Bigg)\nn&+&q_2B(2\xi b_2-A)\f{(1+z)^{\f{1}{2}(b_0-B_1-1)}}{(1+z_i)^{-\f{1}{2}(B_1+3b_0-3)}}\nn&-&q_3B(2\xi b_2-A)\f{(1+z)^{\f{1}{2}(b_0+B_1-1)}}{(1+z_i)^{\f{1}{2}(B_1-3b_0+3)}}\Bigg],\label{deltamfinalz}\\
\delta_{{\rm de}}(z)&=&\f{(1+z_i)^{2(1-b_0)}\delta_{{\rm m}}(z_i)}{BB_1b_2(A_1-A)}\Bigg[AB_1\left(\xi b_2-\f{A_1}{2}\right)\times\nn&&\!\!\!\!\!\!\!\!\!\!\!\!\Bigg(q_1\f{(1+z)^{\f{1}{2}(b_0+B-1)}}{(1+z_i)^{\f{1}{2}(B-3b_0+3)}}-q_0\f{(1+z)^{\f{1}{2}(b_0-B-1)}}{(1+z_i)^{-\f{1}{2}(B+3b_0-3)}}\Bigg)\nn&+&\f{1}{2}q_2A_1B(2\xi b_2-A)\f{(1+z)^{\f{1}{2}(b_0-B_1-1)}}{(1+z_i)^{-\f{1}{2}(B_1+3b_0-3)}}\nn&-&\f{1}{2}q_3A_1B(2\xi b_2-A)\f{(1+z)^{\f{1}{2}(b_0+B_1-1)}}{(1+z_i)^{\f{1}{2}(B_1-3b_0+3)}}\Bigg],\label{deltadefinalz}
\eea
where
\bea\label{qs}
q_0&=&\alpha_2+\f{1}{2}\left(b_0+B-1\right),~~q_1=\alpha_2+\f{1}{2}\left(b_0-B-1\right),\nn
q_2&=&\alpha_2+\f{1}{2}\left(b_0+B_1-1\right),~~q_3=\alpha_2+\f{1}{2}\left(b_0-B_1-1\right).\nn
\eea
We can also integrate Eq. (\ref{deltamdea}) for $\delta_{{\rm de}}=0$, in order to obtain the behavior of matter perturbations in the absence of {\rm DE} perturbations. By doing so we get 
\bea\label{mpertabsde}
\tilde{\delta}_{\rm m}(z)&=&\f{\delta_{{\rm m}}^{\rm i}}{b_0-2q_4-1}\Bigg[(\alpha_2+b_0-q_4-1)\left(\f{1+z}{1+z_i}\right)^{q_4}\nn&-&(\alpha_2+q_4)\left(\f{1+z}{1+z_i}\right)^{b_0-q_4-1}\Bigg],
\eea
where
\be\label{icsdeltatilde}
q_4=\f{1}{2}\left[b_0-\sqrt{b_0^2-2b_0+4b_1+1}-1\right],
\ee
and use has been made of the initial conditions given in Eq. (\ref{adiainitialras}). The obtained expressions (\ref{deltamfinalz}), (\ref{deltadefinalz}) and (\ref{mpertabsde}) provide a wide class of solutions for density perturbations, depending on the Rastall parameter and {\rm EoS} for {\rm DE}. Firstly, from Eq. (\ref{deltamdea}) we realize that {\rm DE} perturbations can act as a source for matter perturbations in such a way that
%\bea\label{coeffdmperteq} 
%\eta=\f{3(4\gamma-1)\left[3(2\gamma-1)w_{\rm de}+6\gamma-1\right]}{2(3\gamma-1)(6\gamma-1)}\Omega_{\rm de},
%\eea
an overdensity in {\rm DE} component could reduce ($b_2<0$) or enhance ($b_2>0$) matter perturbations. Figure (\ref{fig2}) shows the space parameter constructed out of the pair $(\gamma,w_{\rm de})$ where the allowed regions for positive (shaded region) and negative (gray region) values of $b_2$ coefficient are plotted. Interestingly we observe that not always a {\rm DE} overdensity decreases matter clustering and indeed, for certain values of $(\gamma,w_{\rm de})$ parameters we could have matter clustering for $w_{\rm de}<-1/3$. Such a scenario occurs for the shaded region confined between the blue arrow and the curves {\sf CD} and {\sf EF}. Moreover, an underdensity in {\rm DE} perturbations does not necessarily lead to an overdensity in matter component, and instead, can provide both overdense ($b_2<0$) and underdense ($b_2>0$) regions of matter distribution. We also note that for all the points lying on the red arrow ($\gamma=0$), a {\rm DE} overdensity decreases matter clustering for $w_{\rm de}<-1/3$ and vice versa. This is indeed the GR limit of the theory. In Fig. (\ref{fig3}) we have plotted for the behavior of density contrasts against the redshift. We therefore observe that an underdensity in {\rm DE} component (dashed curve within the upper panel) would enhance the evolution of matter component (solid curve) so that matter perturbations can grow even more than the case in which {\rm DE} perturbations are absent (dot-dashed curve). It is also seen that the {\rm DE} perturbations proceed towards formation of void. The lower panel presents the same scenario for which an overdensity occurs in {\rm DE} component and consequently {\rm DE} structures can also form during the evolution of perturbations. We note that the initial stages of structure formation can be adequately investigated within the linear approximation since the density contrast is small enough to neglect the quadratic terms within the system. However, as time goes by, the density contrast grows and consequently non-linear terms would play a major role in the amplitude of perturbations and the fate of the collapse process. This is the subject of the next section.
\section{Non-linear regime with varying $w_{\rm de}$}\label{nonlinearreg}
\begin{figure}
	\includegraphics[scale=0.20]{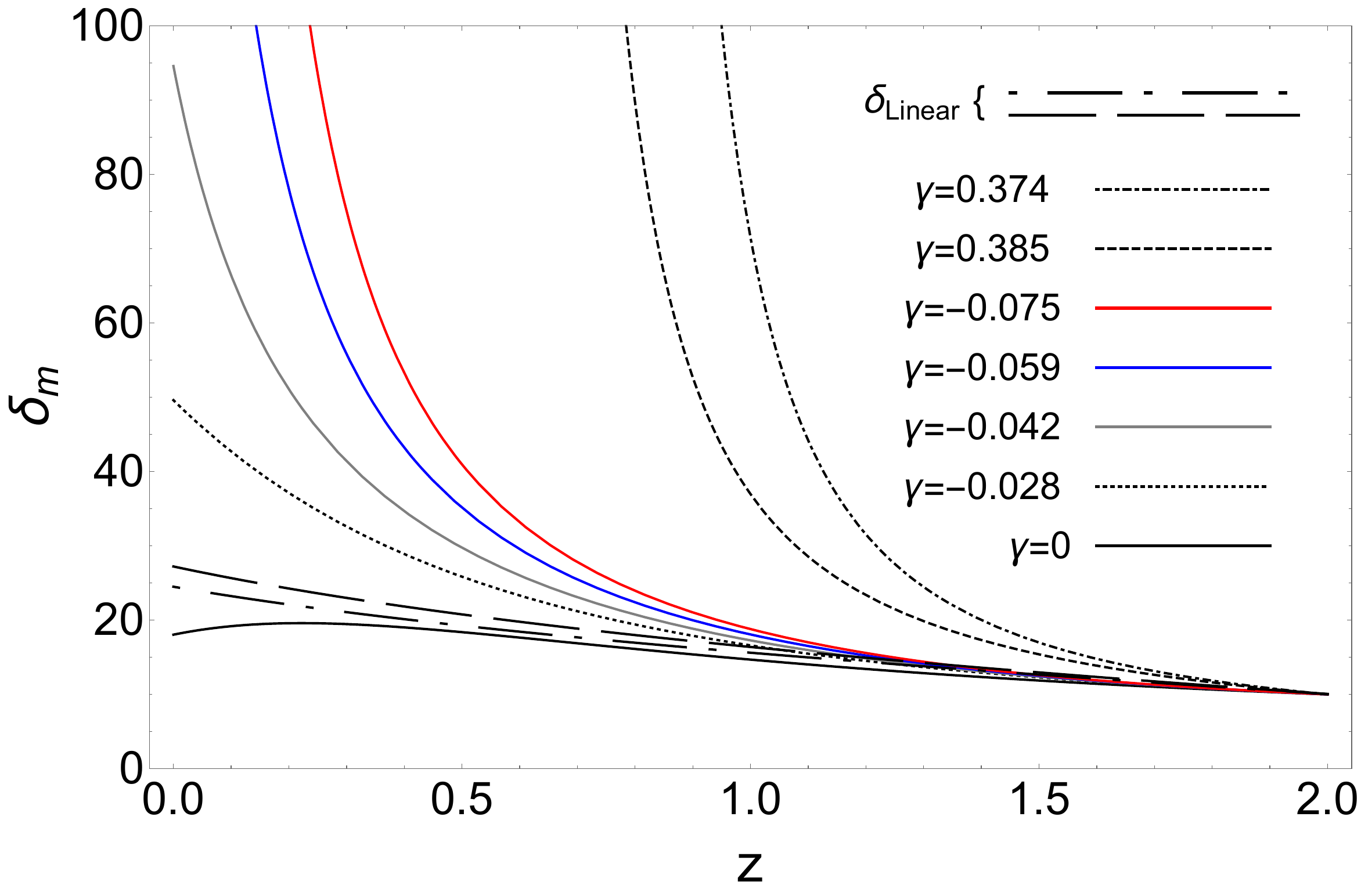}\vspace{0.4cm}
	\includegraphics[scale=0.20]{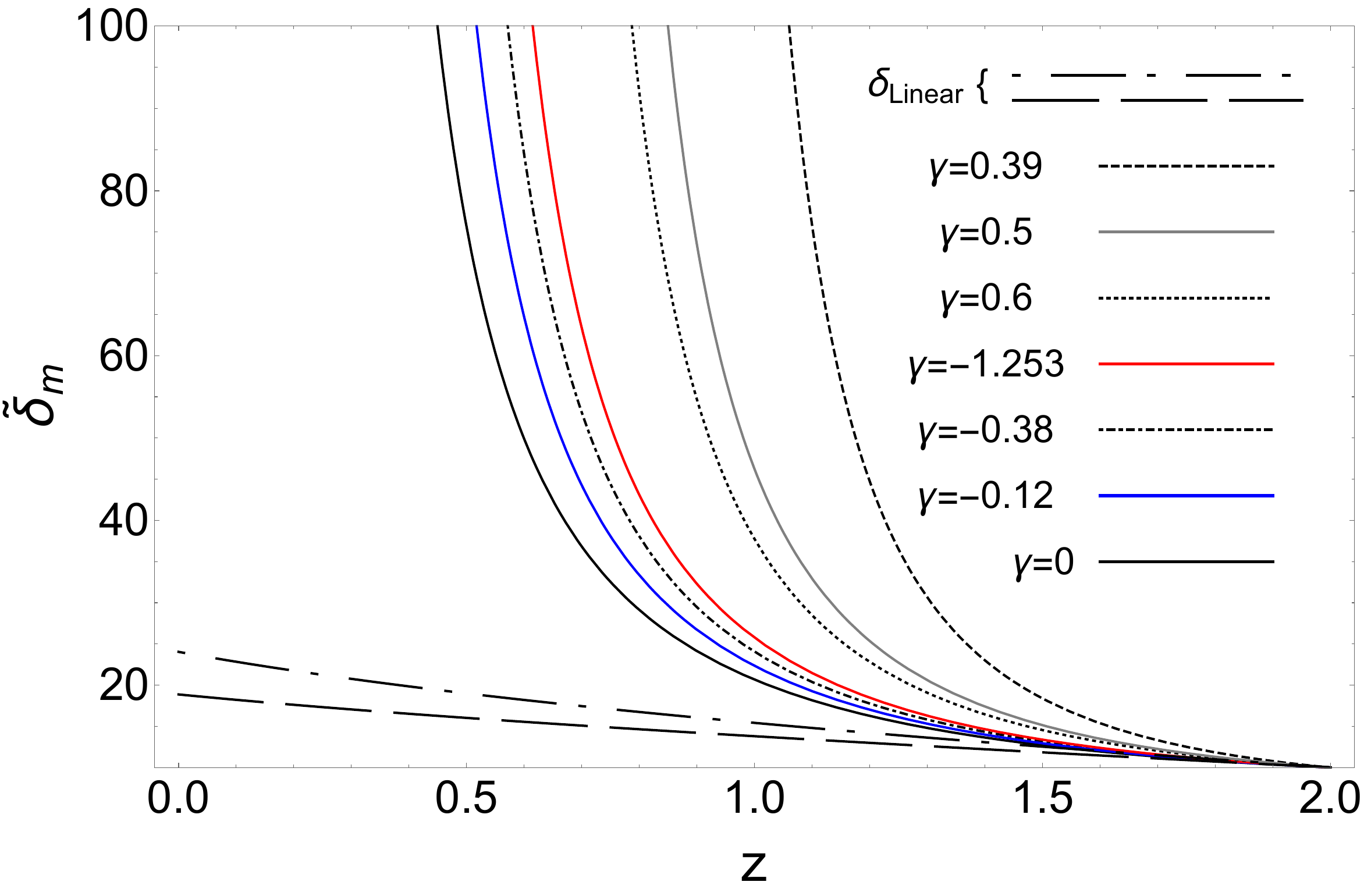}
	\caption{Evolution of the matter perturbations with (upper panel) and without (lower panel) taking into account the {\rm DE} perturbations. The initial value of matter density contrast has been chosen as $\delta_{\rm m}(z=2)=10$. For parametrization of {\rm DE} equation of state we have set $w_0=-0.75$ and $w_1=0.4$. For density parameters we have set $\Omega_{\rm de}=1-\Omega_{\rm m}=3/4.$ The long dashed and long dot-dashed curves represent the evolution of matter perturbations in linear regime for the same values of model parameters as chosen in Fig.(\ref{fig3}).}\label{fig4}
\end{figure}

\begin{figure}
	\includegraphics[scale=0.20]{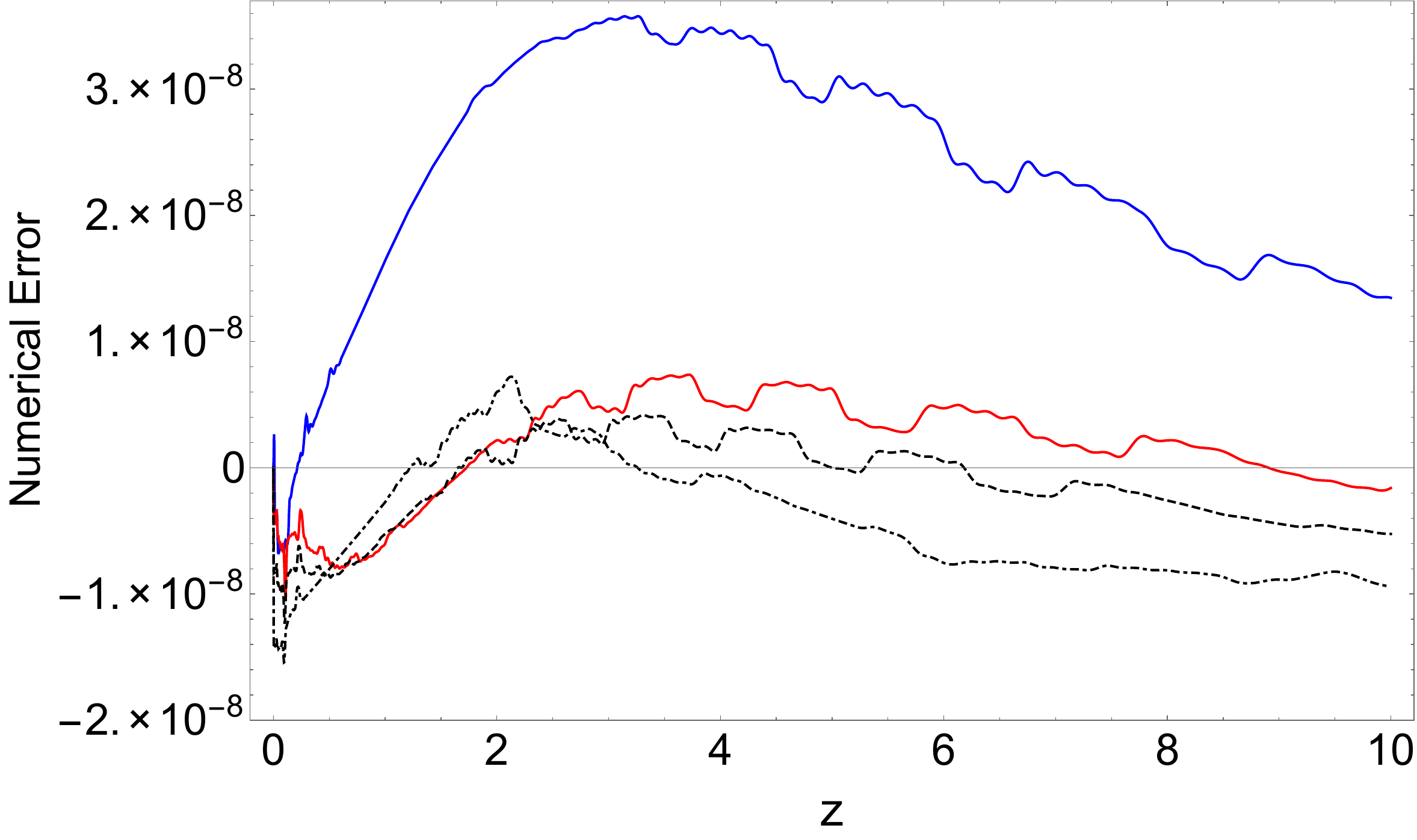}
	\caption{Numerical error associated to the numerical solution given in Fig.(\ref{fig4}). The blue and red curves have been plotted for $\gamma=-0.059$ and $\gamma=-0.075$ and the dashed and dot-dashed curves have been plotted for $\gamma=0.385$ and $\gamma=0.374$, respectively.}\label{fig5}
\end{figure}

When considering the non-linear regime, things get slightly more complicated. Our aim is again to determine the density contrast, however, now we take into account the nonlinear terms within equations (\ref{denconmatter}) and (\ref{dencondark}). By doing so, these equations for a varying {\rm DE} state parameter then read
\bea\label{dmpertnonlin}
a^2\delta^{\prime\prime}_{{\rm m}}&+&ab_0\delta^{\prime}_{{\rm m}}-b_1(1+\delta_{{\rm m}})\delta_{{\rm m}}-b_2(1+\delta_{{\rm m}})\delta_{{\rm de}}\nn&-& a^2b_5\f{\delta_{{\rm m}}^{\prime2}}{1+\delta_{{\rm m}}}=0,
\eea
\bea\label{depertnonlin}
a^2\delta_{{\rm de}}^{\prime\prime}&+&ab_0\delta_{{\rm de}}^{\prime}-b_3(1+\delta_{{\rm de}})\delta_{{\rm m}}-b_4(1+\delta_{\rm de})\delta_{\rm de}\nn&+&\f{a^2\delta_{\rm de}^\prime w_{\rm de}^\prime}{(1+w_{\rm de})(3\gamma(1+w_{\rm de})-1)}-a^2b_6\f{\delta_{\rm de}^{\prime2}}{1+\delta_{\rm de}}=0,\nn
\eea
where 
\bea\label{b56}
b_5=\f{15\gamma-4}{3(4\gamma-1)},~~b_6=\f{3(1+w_{\rm de})(5\gamma-1)-1}{3(1+w_{\rm de})(4\gamma-1)}.
\eea
The above equations can be re-expressed in terms of the redshift using the following relations
\bea\label{ddaddz}
\delta^\prime&=&-(1+z)^2\f{d\delta}{dz},~w_{\rm de}^\prime=-(1+z)^2\f{dw_{\rm de}}{dz}\nn\delta^{\prime\prime}&=&-(1+z)^4\f{d^2\delta}{dz^2}+2(1+z)^3\f{d\delta}{dz}.
\eea
Now if we take the following parametrization for {\rm DE} state parameter~\cite{eosdepar}
\be\label{eosdepar}
w_{\rm de}(z)=w_0+w_1\f{z}{1+z},
\ee
we can solve Eqs. (\ref{dmpertnonlin}) and (\ref{depertnonlin}) using numerical methods. The constants $w_0$ and $w_1$ can be chosen so that they are consistent with observational constraints~\cite{w0w1constraints}. As we expect the numerical solution for non-linear regime is different and depends crucially on the model parameters and initial data. We choose the initial density contrast for matter component to be a finite value at $z=2$. Therefore, the formation of matter structures commences at this redshift and evolves, along with the evolution of {\rm DE} perturbations, until the present time ($z=0$). We find that in response to non-linear perturbations in {\rm DE}, matter perturbations grow at a faster rate and reach a bigger amplitude than expected in linear regime. In Fig. (\ref{fig4}) we have plotted the evolution of DM perturbations in the presence and absence of {\rm DE} perturbations within upper and lower panels, respectively. For $\gamma=0$ (black solid curve), the {\rm DM} density contrast grows monotonically and reaches a finite value at the present time, while for negative values of Rastall parameter, the matter perturbations grow faster so that DM structures may form before reaching the present epoch. For positive values of Rastall parameter (dashed and dot-dashed curves) we observe even more a rate of growth in matter perturbations in such a way that massive objects such as super-clusters can be born within the Universe. As we observe in the lower panel, matter perturbations start to grow from their initial values and diverge as we reach the present time. However, the rate of growth in density contrast for $\gamma=0$ (black solid curve) is lesser than the case in which the Rastall parameter is nonzero. Hence, in comparison to {\rm GR}, we could have massive structures that form faster in Rastall gravity. We also observe that the overall growth rate of matter perturbations in the linear regime (long dashed and long dot-dashed curves) is much slower than the non-linear one. Though, at initial stages of the collapse process, the evolution of density contrasts in both linear and non-linear regimes coincide, non-linear perturbations start detaching from the linear ones as the collapse proceeds to lower redshifts. We also note that as numerical simulations are approximate solutions of the governing equations, differences between numerical solutions and exact ones are expected. The difference is the numerical error. Figure (\ref{fig5}) shows the numerical error associated with the solution of the system (\ref{dmpertnonlin}) and (\ref{depertnonlin}) where we observe that the numerical solution presented in Fig. (\ref{fig4}) satisfies these equations with the accuracy of the order of $10^{-8}$ or less.
\section{Concluding Remarks}\label{conclrems}
In the present work we studied the evolution of {\rm DM} and {\rm DE} perturbations in the context of Rastall theory. In order to
simplify the analysis, we restricted ourselves to spherically symmetric perturbations. Thus, for a spherically symmetric top-hat collapse, we investigated dynamics of density contrast for dark components in both linear and non-linear regimes. In the linear regime, we observed that the Rastall parameter could play an important role in the growth of {\rm DM} perturbations. Moreover, {\rm DM} perturbations could in turn provide a setting for enhancing or decreasing the growth of {\rm DE} perturbations. In {\rm DE} dominated era, we obtained exact solutions for the set of differential equations that govern the dynamics of density contrasts of {\rm DE} and {\rm DM}. We found that {\rm DE} perturbations could increase the rate of growth of {\rm DM} perturbations so that they grow even faster than the case in which {\rm DE} perturbations are absent. Numerical solutions to perturbation equations in non-linear regime show a different scenario. Matter perturbations could grow more rapidly compared with the linear case so that structures which collapse in this manner could be denser than those of linear regime. Hence, the collapse process in non-linear regime could lead to the formation of super clusters of {\rm DM}. We further note that depending on the values and signs of the pair $(\gamma,w_{\rm de})$ other types of solutions can be obtained, in which, {\rm DE} and {\rm DM} perturbations experience oscillations with different frequencies and amplitudes. One then can interpret such a behavior as the ability of spacetime and matter fields to couple to each other in non-minimal way, that the representative of which is the Rastall parameter. We therefore observe that the {\rm DE} and {\rm DM} perturbations in Rastall theory could lead to different fates in comparison to {\rm GR}. From observational viewpoint, Batista et al.~\cite{batista2013} used the data of type Ia supernovae to constrain the Rastall parameter. Their results show that the credible regions in $(\gamma,w_{\rm de})$ plane allow the $\gamma$ parameter to extend to very small and very large values. In this regard, the results of~\cite{batista2013} confirm the allowed values of $(\gamma,w_{\rm de})$ parameters we obtained in Figs. (\ref{FIGWG1}) and (\ref{fig2}). Other observational constraints on Rastall theory have been reported in the literature, for example, the Bayesian method carried out to fit the rotation curves of 16 low surface brightness spiral galaxies, showed that the Rastall parameter is of order $10^{-1}$~\cite{Meirong2020}, which is consistent with the strong lensing estimation~\cite{mnras2019ras}. Also, the study of interiors of neutron stars with realistic {\rm EoS} has reported an astrophysical constraint on this parameter to be of order $10^{-2}$~\cite{oliveira2015}. Despite the successes of Rastall theory in describing cosmological as well as astrophysical scenarios, providing a conclusive constraint on Rastall parameter is still under debate and the model needs to be meticulously confronted with the observational data. From this point of view, though in the present work we did not perform a comprehensive and thorough study of the observational aspects of Rastall theory, we tried to shed some light on the non-equivalence of Rastall gravity and {\rm GR}. A more profound analysis along this line is a subject of our future research work to further put to the test the viability of Rastall gravitation theory in comparison to {\rm GR} with the help of upcoming precise observational data.
\par
As the final remarks, we would like to mention that, though the SC model is an approximation to more realistic collapse scenarios, it provides a suitable framework to investigate the time scale of halo collapse and has proven to be a very useful tool in developing approximate statistical models for the formation and evolution of halos and their abundances. Moreover, this model is very successful in reproducing results of N-body simulations when mass is combined with the function formalism~\cite{DPopolo,DPopolo1}, either in usual minimally coupled DE models~\cite{Pace2010} or in non-minimally coupled DE models~\cite{Pace2014}. However, it is important to extend the basic formalism of SC model in order to incorporate additional terms and make it more realistic. Work along this line has been done e.g., by relaxing the assumption of spherical symmetry~\cite{Hoffman1986,Hoffman1989,Zaroubi1993}, introducing radial motions and angular momentum~\cite{Ryden1987,DPopolo2,Lentz2016,Cupani2011}, studying the effects of DE inhomogeneities~\cite{nunes2006,abramo2009,manera2006,Pace2013,Pereira2010} and shear and rotation~\cite{Reischke2018,DelPolo2018,DelPolo2020,FPace2014,DPopolo3}, see also~\cite{DelPopoloAA,Daichi2016,Pace2017} and references therein. In this regard, extending the results of the present work to include more realistic scenarios could provide us a guideline for better understanding the mutual matter-geometry interaction and possibly its footprints in the formation of cosmic structures.
\end{document}